\input{aipcheck}

\documentclass[
  ,draft            
  ]
  {aipproc}

\layoutstyle{6x9}

\begin{document}

\title{Topological Charge and the Laminar\\ Structure of the QCD Vacuum}

\classification{
                \texttt{11.15.Ha, 11.30.Rd, 12.38.Gc, 12.39.Fe}}

\author{H. B. Thacker}{
  address={University of Virginia, Charlottesville, VA 22904\\}
}

%

\begin{abstract}
Monte Carlo studies of pure glue $SU(3)$ gauge theory using the overlap-based topological
charge operator have revealed a laminar structure in the QCD vacuum consisting of extended, 
thin, coherent, locally 3-dimensional sheets of topological charge embedded in 4D space, 
with opposite sign sheets interleaved. In this talk I discuss the interpretation of these Monte
Carlo results in terms of our current theoretical understanding of theta-dependence and topological
structure in asymptotically free gauge theories. 
\end{abstract}

\maketitle



Recent studies of topological charge using the overlap-based topological charge density
operator on the lattice have
revealed a type of long-range structure that is profoundly different from what might have been
expected in an instanton-based
model of the QCD vacuum. These studies \cite{Horvath03} (see also \cite{Ilgenfritz}) produced the surprising result
that the $q(x)$ distribution
in a typical Monte Carlo gauge configuration is dominated by extended, coherent, thin 3-dimensional sheets
of topological charge. In each configuration, sheets of opposite sign are juxtaposed and are everywhere close together 
in what can roughly be described as a dipole layer which is spread throughout the 4-dimensional Euclidean
space (with various folds and wrinkles). The vacuum is thus permeated with what is locally a laminar structure
consisting of alternating sign sheets or membranes of topological charge. The thickness of these membranes is
typically a few lattice spacings, independent of the physical mass scale, and thus the membranes apparently
become infinitely thin in the continuum limit. This kind of ``subdimensional'' ordering, 
where coherence takes place on manifolds of lower dimensionality than spacetime itself, is closely
related to the appearance of a contact term in the two-point topological charge correlator.
The Euclidean correlator $G(x) =\langle q(x)q(0)\rangle$ {\it is required by spectral considerations to be negative
for any nonzero separation} $|x|>0$. In practice (i.e. in Monte Carlo calculations), this requirement
places severe restrictions on what type of topological charge fluctuations can be dominant. 
For example, the negativity of the correlator rules out the dominance of bulk-coherent lumps of topological charge (e.g. finite
size instantons), as this would lead to a positive correlator over distances smaller than the instanton
size. Since the topological susceptibility can be obtained by integrating the 2-point correlator 
over all $x$, a positive susceptibility can only arise from a delta-function contact term at the
origin. The observed arrangement of thin, nearby layers of $q(x)$ with opposite sign builds up
a positive contact term at $x=0$ while maintaining the required negativity of the correlator for finite separation. 
The range of this contact term on the lattice is associated with the
thickness of the sheets, both of these being a few lattice spacings and 
approaching zero in physical units. This thickness is consistent with the range of non-ultralocality of the overlap
Dirac operator.

Since the initial discovery of coherent topological charge sheets in QCD, similar methods have been
applied to the study of 2-dimensional $CP^{N-1}$ sigma models \cite{Ahmad05}. For $N>3$, the
topological charge distribution was found to be dominated by thin 1-dimensionally coherent membrane-like
structures with interleaved membranes of opposite sign. 
This interleaved arrangement is exactly what one would expect
as the analog of the observed 3-dimensional structures in 4D gauge theory. In both cases, 
the coherent structure has codimension 1, i.e. the dimensionality of a domain wall.  

To interpret these Monte Carlo results, we recall the arguments of Luscher \cite{Luscher78} that nonzero topological
susceptibility implies the presence of a massless pole in the two-point correlator of the Chern-Simons current.
Let us define the {\it abelian} 3-index Chern-Simons tensor
\begin{equation}
\label{eq:CStensor}
A_{\mu\nu\rho} = -Tr\left(B_{\mu}B_{\nu}B_{\rho}+\frac{3}{2}B_{[\mu}\partial_{\nu}B_{\rho]}\right)
\end{equation}
where $B_{\mu}$ is the Yang-Mills gauge potential. We consider the Chern-Simons current that is dual to this tensor,
\begin{equation}
\label{eq:CS4d}
j_{\mu}^{CS} = \epsilon_{\mu\nu\rho\sigma}A_{\nu\rho\sigma}\,\,.
\end{equation}
Although $j_{\mu}^{CS}$ is not gauge invariant, its divergence is the gauge invariant topological charge density
\begin{equation}
\label{eq:csdiv}
\partial_{\mu}j_{\mu}^{CS} = Tr F\tilde{F} = 32\pi^2 q(x) \,\,.
\end{equation}
Choosing a covariant gauge, $\partial_{\mu}A_{\mu\nu\rho}=0$, the correlator of two Chern-Simons currents
has the form
\begin{equation}
\label{eq:cscorr}
\langle j_{\mu}^{CS}(x)j_{\nu}^{CS}(0)\rangle = \int \frac{d^4p}{(2\pi)^4}\; e^{-ip\cdot x}\; \frac{p_{\mu}p_{\nu}}{p^2} G(p^2) \,\,.
\end{equation}
From (\ref{eq:csdiv}) we see that $G(p^2)$ must have a $p^2=0$ pole whose residue is the topological susceptibility,
\begin{equation}
G(p^2) \sim \frac{\chi_t}{p^2} \,\,.
\end{equation}
This long-range correlation constitutes a ``secret long-range order'' of
gauge fields associated with their topological charge fluctuations. Since the CS
current is not gauge invariant, the presence of a $q^2=0$ pole does not imply the existence of a massless particle.
On the other hand, the pole has a gauge invariant residue ($\propto\chi_t$) and cannot be transformed away.
So it characterizes a physically significant long range coherence in the gauge field associated with 
topological charge fluctuations.

Luscher's analysis of QCD topological structure in terms of Wilson bags can be understood as a generalization of the analysis 
of similar properties in the 2-dimensional $CP^{N-1}$ sigma models.
These models provide a quite detailed 2D analog of the coherent structure observed in 4D QCD. 
The $CP^{N-1}$ models have a
$U(1)$ gauge invariance and have classical instanton solutions which come in all sizes. (Just like pure-glue QCD these models 
are classically scale invariant and acquire a mass scale via a conformal anomaly.) 
In 2D U(1) gauge theories like $CP^{N-1}$,
the topological charge density in the continuum is just $(1/2\pi)\epsilon_{\mu\nu}\partial_{\mu}A_{\nu}$ and the 
Chern-Simons current is just the dual of the gauge potential,
\begin{equation}
\label{eq:CS2d}
j_{\mu}^{CS}= \frac{1}{2\pi}\epsilon_{\mu\nu}A_{\nu}
\end{equation}
Just as in 4D QCD, nonzero topological susceptibility implies the presence of a $q^2=0$ pole in the CS current correlator.
But in the two-dimensional case, this same pole appears in the $A_{\mu}$ correlator and
is responsible for confinement of $U(1)$ charge
via a linear coulomb potential.
Thus, in 2-dimensional $U(1)$ theories, topological susceptibility and confinement of $U(1)$ charge 
are equivalent phenomena. An instructive way to illustrate this is to introduce a nonzero $\theta$ term 
in the action over a two-volume $V$ enclosed by a boundary 
$C=\partial V$ with $\theta=0$ outside the boundary. After integration by parts, the theta term 
is equivalent to a Wilson loop around the boundary carrying a charge $\theta/2\pi$:
\begin{equation}
\label{eq:thetaterm}
\exp\left[\frac{i}{2\pi} \int d^2x \theta(x)\epsilon_{\mu\nu}F^{\mu\nu}\right] = \exp\left[\frac{i\theta}{2\pi}\oint_{C} A\cdot dx\right]
\end{equation}
If the topological susceptibility is nonzero 
\begin{equation}
\chi_t=\frac{\partial^2E(\theta)}{\partial \theta^2}|_{\theta=0} >0
\end{equation}
then for small nonzero $\theta$, the vacuum energy density $E(\theta)$ inside the Wilson loop will be greater than that
outside the loop, so it will obey an area law,
\begin{equation}
\langle W(C)\rangle \propto \exp\left[-\left(E(\theta)-E(0)\right)V\right]
\end{equation}
In a Hamiltonian framework, this corresponds to applying a background electric field $\theta$ by putting
opposite charges at either end of the 1-dimensional spatial box. The topological susceptibility is just 
the vacuum polarizability with respect to this field. Periodicity in $\theta\rightarrow \theta+2\pi$ arises
via a discontinuous process of charged pair production at $\theta=\pi$, which screens a unit of electric flux. 
At $\theta=2\pi$ the confining force is completely screened.
This is essentially Coleman's original interpretation of $\theta$-dependence
in the massive Schwinger model \cite{Coleman76}.

In specifying the analogy between 2D U(1) theories and 4D SU(N) gauge theories, 
we take the Chern-Simons currents (\ref{eq:CS4d}) and (\ref{eq:CS2d}) to be directly
analogous. This means that the gauge field $A_{\mu}$
in the 2D theory should be identified {\it not} with the 4-dimensional gauge field, but with the abelian 3-index Chern-Simons
tensor (\ref{eq:CStensor}).
Like the gauge field $A_{\mu}$ in 2 dimensions, this is dual to the Chern-Simons current (\ref{eq:CS4d}).
Similarly, the Wilson loop or line excitations in the 2-dimensional $U(1)$ models correspond not to Wilson loops in 4D, 
but to ``Wilson bags,'' i.e. integrals of the Chern-Simons tensor over a 3-surface $\Sigma$.
\begin{equation}
\label{eq:WB}
B(\Sigma) = \exp\left[i(\theta/2\pi)\int_{\Sigma}A_{\mu\nu\lambda}dx_{\mu}dx_{\nu}dx_{\lambda}\right]
\end{equation}
This is the analog of a Wilson loop in 2D $U(1)$ in the sense that, if $\Sigma$ is a closed 3-surface that forms the 
boundary of a 4-dimensional volume $V$, inserting the Wilson bag
factor (\ref{eq:WB}) in the gauge field path integral is equivalent to including a $\theta$-term in the gauge
action over the 4-volume $V$. The discussion of what happens as we vary $\theta$ from 0 to $2\pi$ is 
completely analogous to the screening of the 2D Wilson loop. For a fractional bag charge $\theta/2\pi$, 
with $0<\theta<2\pi$, the vacuum inside the bag will
have a higher energy than the $\theta=0$ vacuum outside. The expectation of the 
Wilson bag integral thus satisfies a volume law analogous to the area law for the Wilson loop in 2D, and
there will be ``bag confinement,'' 
a confining force between the walls of a fractionally charged bag. 
At $\theta=2\pi$, the confining force between bag walls disappears. Integer charged bag surfaces are thus
free to percolate throughout the vacuum. The topological
charge is the curl of the Chern-Simons tensor, so for a uniform $A_{\mu\nu\lambda}$ which is nonzero on a 
flat bag surface, the topological charge distribution is a dipole layer consisting of thin, coherent positive and 
negative layers on either side of the bag surface. Like the Wilson line excitations in the 
$CP^{N-1}$ models \cite{Ahmad05}, a vacuum full of unit-charged Wilson bags provides a reasonable model for interpreting
the topological charge structure observed in lattice Monte Carlo simulations.

The role of Wilson bag excitations in QCD is greatly illuminated by string/gauge duality, exploiting 
the dual relationship between topological charge in gauge theory and
Ramond-Ramond charge in IIA string theory \cite{lat06}. 
As Witten showed \cite{Witten98}, the string/gauge correspondence 
nicely confirms the k-vacuum/Wilson bag/domain wall scenario originally suggested by
large-$N_c$ chiral Lagrangian arguments \cite{Witten80}.
It also points to the correct candidate for the
string theory analog of a Wilson bag in gauge theory.
In IIA string theory, the Wilson bag can be interpreted as the
holographic image of a D6-brane (which is wrapped around a compact $S_4$ and therefore
appears as a 2-brane or membrane in 3+1 dimensions). In particular, the defining
property of the Wilson bag, namely that the value of $\theta$ jumps by
$\pm 2\pi$ when crossing the surface, is in fact nothing but the statement of quantization
of Ramond-Ramond charge on a D6-brane. 

This work was supported in part by the Department of Energy under grant DE-FG02-97ER41027.







\bibliographystyle{aipproc}   

\bibliography{sample}

\IfFileExists{\jobname.bbl}{}
 {\typeout{}
  \typeout{******************************************}
  \typeout{** Please run "bibtex \jobname" to optain}
  \typeout{** the bibliography and then re-run LaTeX}
  \typeout{** twice to fix the references!}
  \typeout{******************************************}
  \typeout{}
 }


\end{document}